%% file: 0main.tex
\documentclass[letterpaper,twocolumn,10pt]{article}

\usepackage{usenix}

\usepackage{tikz}
\usepackage{amsmath}

\usepackage{filecontents}
\usepackage{xspace}
\usepackage{url} 
\usepackage[altpo]{backnaur}
\usepackage{booktabs, multirow} %
\usepackage[ruled, vlined, linesnumbered, commentsnumbered, longend]{algorithm2e}
\input{macros}
\usepackage{subcaption}

\usepackage{xspace}
\usepackage{enumitem}
\usepackage{booktabs, multirow} %
\usepackage{soul}
\usepackage{algorithmic}
\usepackage[ruled, vlined, linesnumbered, commentsnumbered, longend]{algorithm2e}
\usepackage{graphicx}
\usepackage{tcolorbox}
\usepackage{wrapfig}
\usepackage[normalem]{ulem}
\usepackage{subcaption}
\usepackage{xcolor}
\usepackage{listings}
\usepackage{balance}
\usepackage{setspace}
\usepackage{mathtools}
\usepackage{makecell}
\usepackage{amsthm}
\usepackage{tikz}
\usepackage{graphicx,calc, scalerel}
\usepackage{caption}
\usepackage{bbm}
\usepackage{pifont}
\usepackage{pdfpages}
\usepackage{tabularx}
\usepackage{xurl}
\usepackage{hyperref}

\begin{document}

\title{May the Feedback Be with You! Breaking the Seal of Feedback-Driven \\ Deep Learning Framework Fuzzing via Large Language Models}

\author{
{\rm Shaoyu Yang} \\ 
Nanjing University \\ 
shaoyuyoung@gmail.com
\and 
{\rm Chunrong Fang} \\ 
Nanjing University \\ 
fangchunrong@nju.edu.cn
\and 
{\rm Haifeng Lin} \\ 
Nanjing University \\ 
linhaifeng0716@163.com
\and
{\rm Xiang Chen} \\ 
Nantong University \\ 
xchencs@ntu.edu.cn
\and
{\rm Jia Liu} \\ 
Nanjing University \\ 
liujia@nju.edu.cn
\and 
{\rm Zhenyu Chen} \\ 
Nanjing University \\ 
zychen@nju.edu.cn
}
\maketitle

\begin{abstract}
Deep Learning (DL) frameworks have served as fundamental components in DL systems over the last decade. However, bugs in DL frameworks could lead to catastrophic consequences in critical scenarios. A simple yet effective way to find bugs in DL frameworks is fuzz testing (Fuzzing). Existing approaches focus on test generation, leaving execution results with high semantic value (\eg coverage information, bug reports, and exception logs) in the wild, which can serve as multiple types of feedback.

To fill this gap, we propose \tactic to effectively utilize the feedback information, which comprises two Large Language Models (LLMs): analysis LLM and generation LLM. Specifically, analysis LLM infers analysis summaries from feedback information, while the generation LLM creates tests guided by these summaries. Furthermore, based on multiple feedback guidance, we design two additional components: \textit{(i)} a feedback-aware simulated annealing algorithm to select operators for test generation, enriching test diversity. \textit{(ii)} a program self-repair strategy to automatically repair invalid tests, enhancing test validity. We evaluate \tactic on the two most popular DL frameworks, and experiment results show that \tactic can improve line code coverage of \pt and \tf by \revise{4.48\% and 9.14\%} over four state-of-the-art baselines. By the time of submission, \tactic has detected 104 previously unknown bugs for \pt and \tf, with 93 confirmed as new bugs, 53 already fixed. \revise{14 vulnerabilities have been assigned CVE IDs, among which 7 are rated as high-severity with a CVSS score of "7.5 HIGH".}
\end{abstract}


\input{1intro}

\input{2background}

\input{3approach}

\input{4implementation}

\input{5evaluation}

\input{6discussion}

\input{7related}

\input{8conclusion}

\section*{Acknowledgment}
We thank Wei Cheng, Yanzhou Mu, and Zhiyuan Li for their insightful feedback on the manuscript.

\bibliographystyle{plain}
\bibliography{refs}

\appendix
\section{Failure Case Analysis}
\begin{figure}[htbp]
	\centering
	\includegraphics[width=0.85\columnwidth]{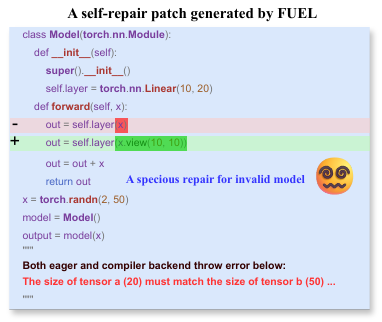}
	\caption{A failed self-repair of \tactic}
	\label{fig:failure}
\end{figure} 
As discussed in our ablation study, \tactic can repair most of the invalid models. However, not all the invalid models can be repaired. After our manual check, we find that failed repair attempts typically involve programs with excessive operators and complex constraints, where LLMs erroneously apply shape-conversion operators (\eg \CodeIn{View} and \CodeIn{Unsqueeze}), zooming shape errors. Figure~\ref{fig:failure} presents a representative failure mode involving operator compatibility. The original model defines a \CodeIn{Linar} operator expecting an input dimension of \CodeIn{10}, but receives a tensor \CodeIn{x} with shape \CodeIn{(2, 50)}. Triggered by the dimension mismatch error, \tactic attempts to repair the invalid model by reshaping \CodeIn{x} into \CodeIn{(10, 10)} using additional \CodeIn{.view()}. While this satisfies the \CodeIn{Linear} operator's constraint (producing an output shape \CodeIn{(10, 20)}), it implicitly destroys the batch dimension semantics. This ``specious'' fix causes a crash in the subsequent residual connection \CodeIn{out + x}, where the shapes \CodeIn{(10, 20)} and \CodeIn{(2, 50)} become irreconcilable. Finally, the patched model is still invalid, and both backends throw \CodeIn{RuntimeErrors}. A promising future direction is that LLM generates shape-agnostic DL programs (\ie fuzz drivers) while leveraging constraint-solving techniques (\eg Z3Prover~\cite{z3-solver}) to automatically resolve tensor shape and API parameters, thereby producing valid and diverse DL models through hybrid neuro-symbolic approaches.

\end{document}

%% file: macros.tex
\newcommand{\ignore}[1]{}

\definecolor{applegreen}{rgb}{0.55, 0.71, 0.0}

\newcommand{\revise}[1]{{#1}}

\newcommand{\parabf}[1]{\noindent\textbf{#1}}
\newcommand{\CodeIn}[1]{{\small \texttt{#1}}}

\newcommand{\tactic}{\textsc{FUEL}\xspace}
\newcommand{\algo}{FASA\xspace}

\newcommand{\titanfuzz}{\textsc{TitanFuzz}\xspace}
\newcommand{\fuzzgpt}{\textsc{FuzzGPT}\xspace}
\newcommand{\whitefox}{\textsc{WhiteFox}\xspace}
\newcommand{\fuzzall}{\textsc{Fuzz4All}\xspace}

\newcommand{\sedar}{Sedar\xspace}
\newcommand{\nnsmith}{\textsc{NNSmith}\xspace}
\newcommand{\neuri}{\textsc{NeuRI}\xspace}

\newcommand{\woals}{\textsc{Fuel$_{als}^{wo}$}\xspace}
\newcommand{\wosa}{\textsc{Fuel$_{sa}^{wo}$}\xspace}
\newcommand{\wocov}{\textsc{Fuel$_{cov}^{wo}$}\xspace}
\newcommand{\wobug}{\textsc{Fuel$_{bug}^{wo}$}\xspace}
\newcommand{\woexc}{\textsc{Fuel$_{exc}^{wo}$}\xspace}

\newcommand{\genm}{$\mathcal{M_{G}}$\xspace}
\newcommand{\alsm}{$\mathcal{M_{A}}$\xspace}

\newcommand{\pt}{PyTorch\xspace}
\newcommand{\tf}{TensorFlow\xspace}
\newcommand{\ptinductor}{PyTorch Inductor\xspace}
\newcommand{\tfxla}{TensorFlow XLA\xspace}

\newcommand{\eg}{\emph{e.g.,}\xspace}
\newcommand{\ie}{\emph{i.e.,}\xspace}

\definecolor{myblue}{HTML}{99D6FF}
\definecolor{myred}{HTML}{FFCCCC}

%% file: 1intro.tex
\section{Introduction}
\label{sec:intro}

Deep Learning (DL) techniques have achieved success in various domains over the last decade, including healthcare~\cite{healthcare}, chatbots~\cite {chatgpt}, autonomous driving~\cite{huval2015autonomousdriving}, and finance security~\cite{finance}. In these domains, popular DL frameworks like PyTorch~\cite{PyTorch} and TensorFlow~\cite{Tensorflow} help developers build, train, compile, and deploy DL models. Despite their widespread use, these frameworks remain susceptible to bugs, which can pose serious consequences, particularly in safety-critical applications~\cite{uberkill}. Hence, ensuring the quality of DL frameworks is essential, and various automated testing techniques~\cite{wang2020lemon,pham2019cradle,mu2024devmut,liu2023nnsmith} are proposed to detect bugs in DL frameworks. Among them, fuzzing~\cite{SuttonFuzzingBook,herrera2021seed,gong2025snowplow} is a prominent technique that systematically generates numerous inputs to uncover potential bugs. Traditional fuzzers~\cite{wang2020lemon,liu2023nnsmith} typically require manual design for mutation rules or operator specifications, which demands extensive domain knowledge and great human effort.

To alleviate the above issue, recent research has introduced Large Language Models (LLMs)~\cite{chatgpt, openai2023gpt4} into fuzzing, motivated by their success in other fields~\cite{SEsur}. One of the first classical studies is \titanfuzz~\cite{titanfuzz}, whose key insight is that LLMs can implicitly learn both language syntax and intricate DL API constraints for DL program generation. Inspired by \titanfuzz, some follow-up studies~\cite{fuzzgpt,promptfuzz,whitefox,kernelgpt} are proposed to explore the potential of LLMs in fuzzing. The core idea behind these studies is to leverage the strong generative abilities of LLMs to generate or mutate inputs directly. Under such a trend, the ``magnificent edifice'' of fuzzing has already been built, but one dark cloud still hangs in the air: \textbf{\textit{Have we fully exploited the potential of feedback information during the fuzzing loop, especially the semantically rich ones?}} Before answering this question, we first analyze the limitations of existing studies:

\begin{figure}[htbp]
	\centering
 	\includegraphics[width=0.85\columnwidth]{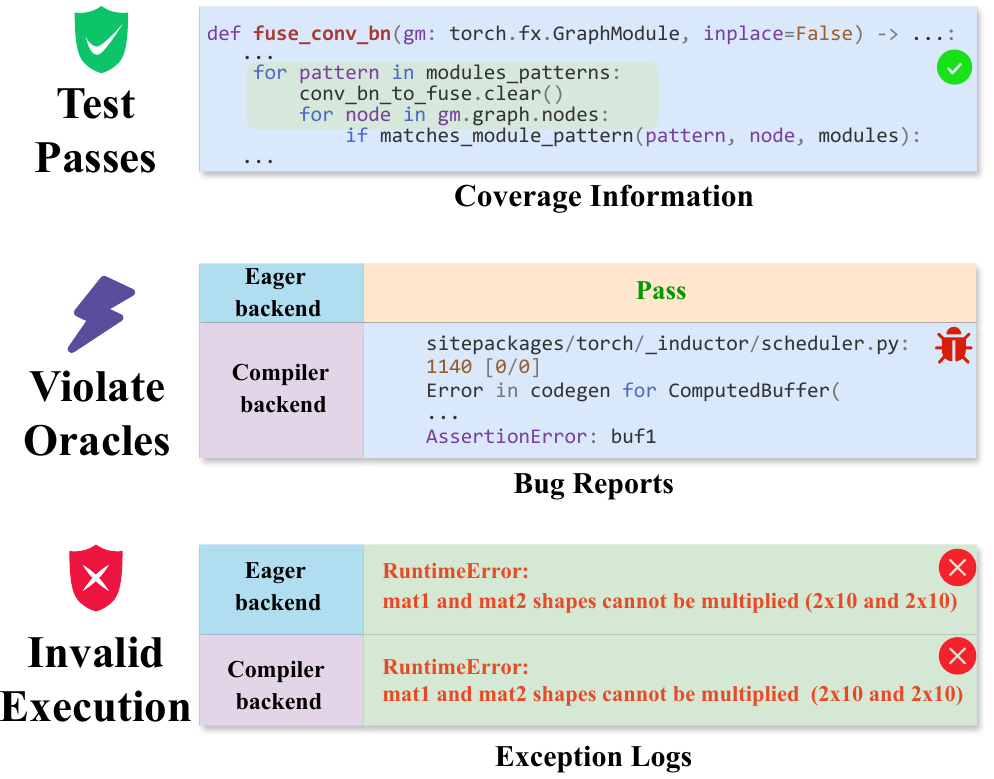}
	\caption{Motivation example of illustrating Limitation \#1\&2}
	\label{fig:motivation1}
\end{figure}

\textbf{Limitation \#1: Overlooking multiple types of feedback.} Previous approaches primarily focus on leveraging the powerful generation capabilities of LLMs to generate random inputs or edge cases. However, they tend to overlook the rich information derived from multiple feedback types. For example, \titanfuzz~\cite{titanfuzz} and \fuzzgpt~\cite{fuzzgpt} do not utilize any feedback in their fuzzing loop, merely employing differential testing~\cite{evans2007difftesting} to identify potential bugs. 
Regrettably, the absence of feedback guidance leads to decreased code coverage and leaves more bugs undetected.
In contrast, PromptFuzz~\cite{promptfuzz} employs a coverage-guided strategy to encourage LLMs to generate more diverse tests, alleviating the coverage plateaus problem. \whitefox~\cite{whitefox} advances this idea by injecting compiler-optimization heuristics into the feedback loop. 
However, these techniques still only utilize a \textit{single} and \textit{simple} type of feedback during the whole fuzzing campaign and fail to consider multiple types of feedback simultaneously. As Figure~\ref{fig:motivation1} shows, each test execution yields three mutually exclusive feedback categories: \textit{coverage information}, \textit{bug reports}, and \textit{exception logs}. However, how to fuse and exploit such complex feedback remains unexplored.

\textbf{Limitation \#2: Coarse-grained feedback information analysis.} Each test execution in fuzzing yields various semantic information, which is usually a mixture of programming language and natural language. 
state-of-the-art (SOTA) feedback-driven fuzzers still distill feedback information into coarse signals (\eg coverage increase), thereby discarding the fine-grained, context-rich semantic information.
Even PromptFuzz~\cite{promptfuzz}, which mitigates coverage plateaus, only focuses on whether new code is covered, ignoring semantics from concrete code snippets. This coarse-grained manner prevents the fuzzer from reasoning about the actual runtime state of the system under test (SUT). As Figure~\ref{fig:motivation1} shows, coverage information pinpoints specific code snippets, bug reports contain symptom descriptions, and exception logs provide detailed stack traces. Each feedback category can be effectively processed by LLMs, which transform raw feedback into actionable insights to guide subsequent test generation.

In addition to the utilization of feedback, existing techniques still have the following limitations in test generation.

\begin{figure}[htbp]
	\centering
	\includegraphics[width=0.85\columnwidth]{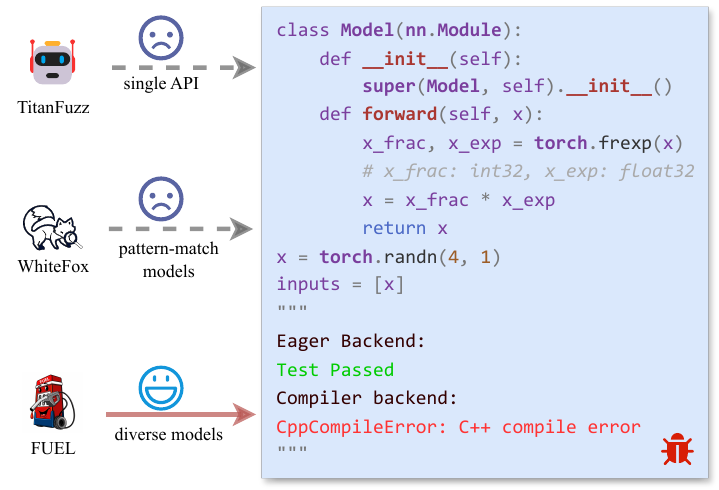}
	\caption{Motivation example of illustrating Limitation \#3}
	\label{fig:motivation2}
\end{figure}  
\textbf{Limitation \#3: Insufficient test diversity.}
\titanfuzz and \fuzzgpt restrict themselves to API-level fuzzing (introduced in \S~\ref{subsec:bg-framework}), exhaustively enumerating isolated operators one by one. Although this paradigm achieves high API coverage, it overlooks bugs that manifest only when multiple operators interact within DL models. Model-level fuzzers such as \whitefox~\cite{whitefox} synthesize tests that target compiler optimization patterns (\eg operator fusion), yet this optimization-centric focus sacrifices test diversity, as many latent bugs lie outside the chosen optimization patterns. For instance, Figure~\ref{fig:motivation2} shows a test in which \CodeIn{torch.frexp} returns two tensors (\CodeIn{torch.int32} and \CodeIn{torch.float32}) which are subsequently element-wise multiplied. Eager backend of \pt silently handles type promotion, whereas the compiler backend crashes with \CodeIn{CppCompileError}. \titanfuzz misses this bug due to their single-API testing paradigm. \whitefox was also unable to detect this bug, as it can not generate such models that mismatch any optimization pattern. A natural and intuitive solution is to enable the LLM to consider more operator combinations by supplying external context information, thereby diversifying the generated tests.

\begin{figure}[htbp]
	\centering
	\includegraphics[width=0.85\columnwidth]{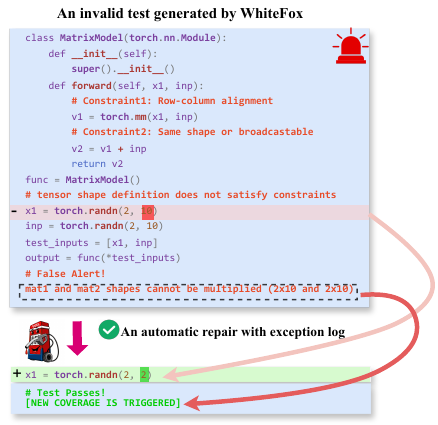}
	\caption{Motivation example of illustrating Limitation \#4}
	\label{fig:motivation3}
\end{figure} 

\textbf{Limitation \#4: Limited validity of generated tests.} 
Even SOTA LLMs currently struggle to create valid tests for DL frameworks~\cite{deng2024can}. Existing techniques prompt LLMs with few-shot examples through in-context learning, but this paradigm poses limitations to LLMs' understanding of the constraints on DL operators. Figure~\ref{fig:motivation3} shows an invalid test generated by \whitefox that combines \CodeIn{torch.mm} and \CodeIn{Add}, imposing two constraints: \textit{(i)} the first input of \CodeIn{torch.mm} must be column-aligned with the second; \textit{(ii)} the inputs of \CodeIn{Add} must be either identically shaped or broadcastable. The generated tensors, however, have shapes $(2 \times 10)$ and $(2 \times 10)$, violating the above constraints and triggering a shape-mismatch error, which is a false positive. Our manual check reveals that \textbf{47.54\% (290/610)} of the tests generated by \whitefox are invalid, similar to the above example. These failures not only waste fuzzing budgets but also lose opportunities to uncover bugs within operator interactions. Notably, we observe that most invalid tests are easy to fix. In this case, simply reshaping \textit{x1} from $(2 \times 10)$ to $(2 \times 2)$ generates a valid test that executes successfully and explores previously uncovered code paths. Equipped with the exception logs, LLMs are capable of automatically conducting such small-scale repairs, converting invalid executions into valid ones, thereby improving code coverage and bug detection capabilities.

\textbf{Insight.} LLMs have the potential to analyze multiple types of feedback information semantically and thereby generate diverse and valid tests. This motivates us to design two LLMs with distinct roles: \textit{(i)} an analysis LLM distills \textit{complicated} and \textit{redundant} feedback information into concise summaries; \textit{(ii)} a generation LLM synthesizes \textit{diverse} and \textit{valid} tests guided by analysis summaries during the fuzzing loop.

\textbf{Our work.} In this study, we propose \tactic, the first feedback-driven fuzzer that fully utilizes feedback information with two LLM-based agents, namely analysis LLM (\alsm) and generation LLM (\genm). The key idea of our design is that LLMs could understand and analyze valuable feedback information. \alsm first automatically infers the analysis summaries from the feedback. Then, analysis summaries are used to guide the \genm to generate tests that help to explore new code coverage or trigger new bugs (addressing \textbf{Limitation \#1} and \textbf{Limitation \#2}). However, tests generated by LLMs usually contain common \textit{well-tested} operators (\eg \CodeIn{Conv2d}, \CodeIn{ReLU}), resulting in the lack of some uncommon \textit{poorly-tested} operators (\eg \CodeIn{ConvTranspose}, \CodeIn{InstanceNorm}). This behavior reduces test \textit{diversity}. To solve this, we collect core operators from the DL framework API document as a retrieval set and design a feedback-aware simulated annealing algorithm to select operators from the retrieval set in each iteration. The retrieved operators, along with the analysis summary, are filled into the prompt of the \genm as context, allowing \genm to generate tests by using more operators (addressing \textbf{Limitation \#3}). To improve test \textit{validity}, \tactic introduces a program self-repair strategy to repair invalid tests. In simple terms, \alsm in \tactic generates the repair strategy based on the exception log, which prompts the \genm to repair the test from the last iteration (addressing \textbf{Limitation \#4}). 

In summary, our work makes the following contributions:
\begin{itemize}[noitemsep, leftmargin=15pt, topsep=2pt]
    \item \textbf{Direction.} To the best of our knowledge, we are the first to leverage LLMs to \textit{semantically} analyze \textit{multiple types} of feedback in fuzzing. Furthermore, beyond DL framework testing, this paradigm is also capable of conducting targeted fuzzing for other large and complex software systems by customizing more types of feedback (\eg execution time and memory usage). Our promising results encourage further exploration in this direction.
    
    \item \textbf{Approach.} We implement \tactic as a practical fuzzer for DL framework testing, which utilizes feedback information and repairs invalid tests. \tactic is available at \url{https://github.com/NJU-iSE/FUEL}.
    
    \item \textbf{Evaluation.} Our experimental results show that the tests generated by \tactic achieve superior performance compared to SOTA baselines (\neuri, \titanfuzz, \nnsmith, and \whitefox), improving line coverage on two DL frameworks (\pt and \tf) by 4.48\% and 9.14\%, respectively. 
    Additionally, ablation studies validate the contribution and effectiveness of each component in \tactic’s design. 

    \item \textbf{Real-World Contribution.} So far, \tactic has detected 104 previously unknown bugs, with 93 already confirmed, 53 already fixed, and 14 assigned CVE IDs. 14 detected bugs were labeled as \textit{high priority}, and one issue was labeled as \textit{utmost priority}.
\end{itemize}

%% file: 2background.tex
\section{Background}
\label{sec:bg}
\subsection{DL Framework Testing}
\label{subsec:bg-framework}
Previous studies on DL framework testing can be classified into API-level fuzzers~\cite{wei2022free,xie2022docter,titanfuzz} and model-level fuzzers~\cite{wang2020lemon,liu2023nnsmith,whitefox}. API-level fuzzers target testing each DL framework API by infilling its parameters. Notably, a DL framework API usually corresponds to a DL operator. For instance, \CodeIn{torch.Conv2D}, a \pt API (corresponding to 2-D convolution operator) can be filled with some parameters by API-level fuzzers, including \CodeIn{kernel\_size}, \CodeIn{stride}, and \CodeIn{padding}. Then, a 4-D tensor will be created as input for \CodeIn{torch.Conv2D}. Finally, comparing outputs across different hardware (\eg CPU and GPU) to detect potential bugs.

However, this paradigm is limited in its ability to detect bugs resulting from the intricate compositions of diverse operators in DL models. Therefore, model-level fuzzers are proposed to synthesize or mutate DL models that encapsulate a wide range of operators. A model-level test typically comprises two components: \textbf{model definition} and \textbf{input tensors}. As shown in Figure~\ref{fig:motivation3}, the model definition is a Python class inheriting from the framework's base \CodeIn{Model} class (\eg \CodeIn{torch.nn.Module}). Meanwhile, the input tensors supply external tensor data to the model (\eg \CodeIn{torch.randn(4, 1)}). Model-level fuzzers therefore strive to generate both the DL model (\ie model definition) and input tensors that satisfy the model constraint, jointly forming a single model-level test. In this work, our designed FUEL is a model-level fuzzer, which aims to create diverse yet valid DL models.

\label{sec:approach}
\begin{figure*}[htbp]
	\centering
	\includegraphics[width=1\textwidth]{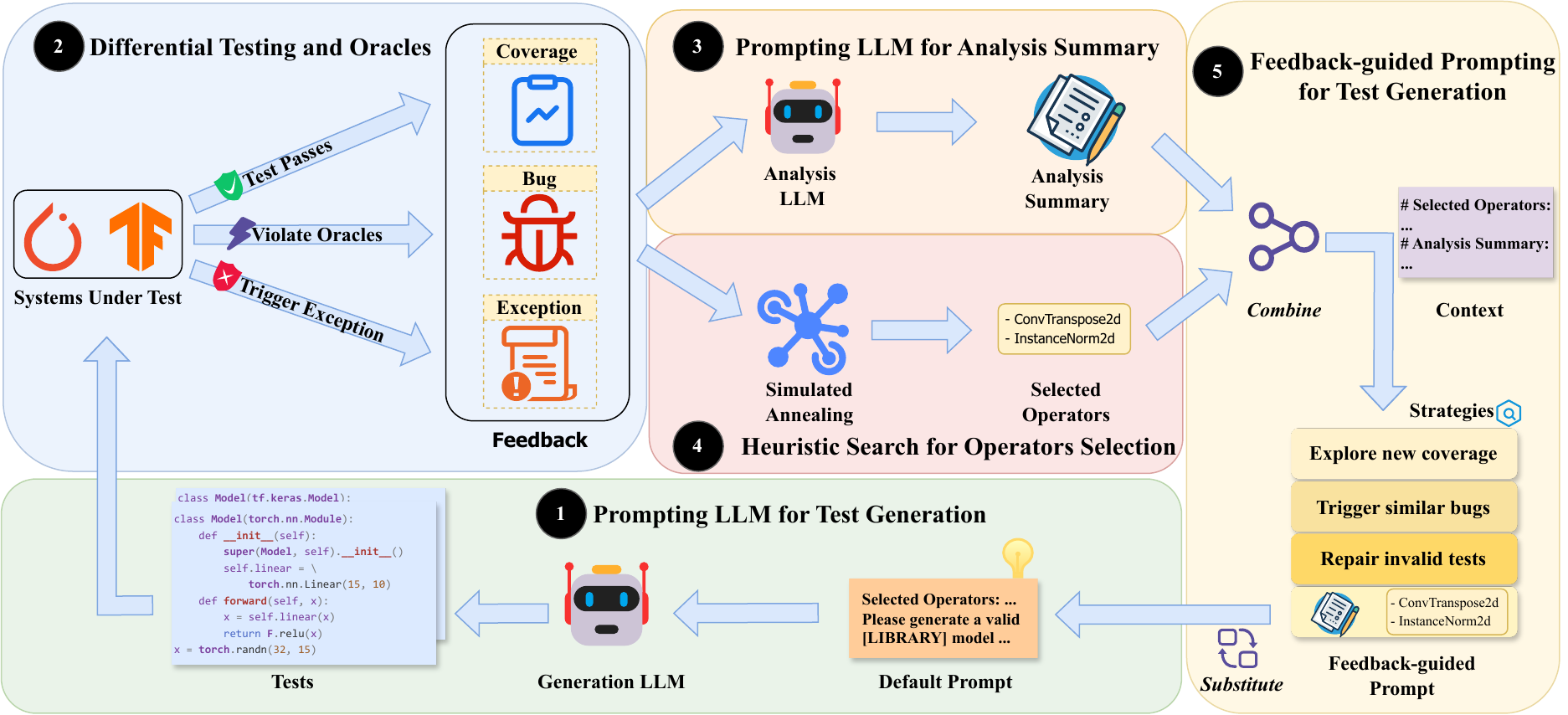}
	\caption{Overview of \tactic}
	\label{fig:overview}
\end{figure*}  

\subsection{LLM-based Fuzzing}
\label{subsec:bg-llm}
Until now, LLMs have achieved remarkable success in code intelligence tasks~\cite{hou2024llm4se}. Modern LLMs, typically based on Transformers, can be categorized into two types: Dense Transformer~\cite{transformer} and Mixture-of-Experts (MoE) architecture~\cite{jiang2024mixtral}. Dense Transformers (\eg Llama-3~\cite{dubey2024llama} and Qwen-2.5-Coder~\cite{hui2024qwen2.5-coder}) employ uniform parameter activation across all layers, enabling full contextual interaction through self-attention mechanisms, while MoE architecture models (\eg Mixtral~\cite{jiang2024mixtral} and DeepSeek-V3~\cite{deepseekai2024deepseekv3}) introduce sparse activation via expert routing algorithms, dynamically allocating computations to specialized subnetworks for improved scalability and efficiency. Recently, reasoning models (\eg DeepSeek-R1~\cite{deepseekai2025deepseekr1}, GPT-o1~\cite{openai-o1}) based on chain-of-thought (CoT)~\cite{wei2022cot} have attracted worldwide attention, showing impressive performance on math- and code-related benchmarks.

Impressive performance of LLMs motivates researchers to bring LLMs into software testing~\cite{wang2024software}, especially fuzzing~\cite{llmfuzzsurvey}. In LLM-based fuzzing, LLM is usually employed as the test generator to synthesize programs. Along with instructions and the few-shot examples in prompts, it directly generates test code (similar to the code snippet in Figure~\ref{fig:motivation3}). Specifically, these approaches primarily focus on using LLMs to generate or mutate inputs based on static prompts. However, they often overlook the execution results of generated tests, failing to utilize such valuable feedback to guide subsequent generation. In this context, \tactic augments the fuzzing loop with an analysis LLM to distill raw feedback information into actionable summaries, explicitly guiding the generation LLM to enrich test diversity and validity.


%% file: 3approach.tex
\section{\tactic Design}

Figure~\ref{fig:overview} depicts the end-to-end workflow of \tactic, integrating a \alsm with a feedback-aware simulated annealing algorithm to utilize feedback information in a fully automated manner and subsequently combining analysis results to guide \genm to generate diverse and valid DL models. We show the details of each phase in the following subsections.

\subsection{Prompting LLM for Test Generation}
While generally employing a feedback-driven fuzzing approach, \tactic is designed to generate tests without feedback guidance in two key scenarios: \textit{(i)} In the initial iteration of fuzzing: \genm performs purely stochastic generation based on predefined prompts, as currently no prior feedback exists for guidance. \textit{(ii)} Upon model repair failure: \tactic incorporates a collaborative self-repair strategy (as introduced in \S~\ref{subsec:feedback} and \S~\ref{subsec:guidance}) that engages both \alsm and \genm. When this self-repair fails in the current iteration, \genm temporarily disregards previous feedback and reverts to a non-guided generation in the next iteration cycle. The motivation behind this strategy is to prevent \tactic from being trapped in an \textit{infinite repair loop} (\ie attempt successive iterations persistently but fail to resolve preceding errors).   

\begin{figure}[htbp]
    \centering
    \captionsetup[subfigure]{justification=centering, aboveskip=3pt, belowskip=2pt}
    \begin{subfigure}{\linewidth}
        \centering
        \includegraphics[width=0.85\linewidth]{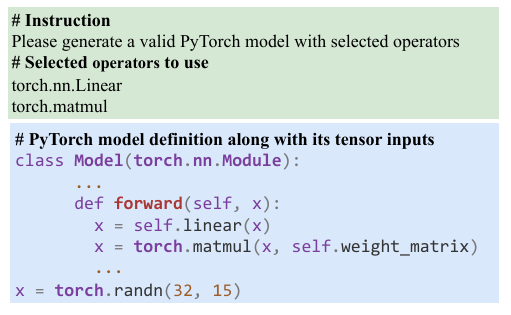}
        \caption{A Few-Shot Example of \pt}
        \label{fig:gen-prompt-default-1}
    \end{subfigure}

    \begin{subfigure}{\linewidth}
        \centering
        \includegraphics[width=0.85\linewidth]{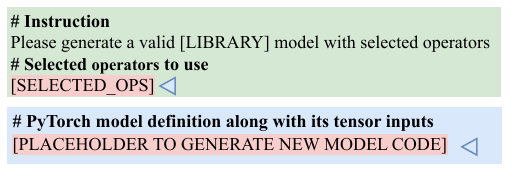}
        \caption{Default Prompt Template}
        \label{fig:gen-prompt-default-2}
    \end{subfigure}
    \captionsetup{aboveskip=5pt}
    \caption{Default prompt template of \tactic}
    \label{fig:gen-prompt-default}
\end{figure}

To accommodate these two scenarios, we design specialized default generation prompts that ensure continuous test generation even in scenarios where feedback is unavailable or repair stagnation occurs. The transition between feedback-driven and default generation is algorithmically controlled to optimize \textit{exploration-exploitation} balance throughout the whole fuzzing campaign. Figure~\ref{fig:gen-prompt-default-1} shows a few-shot example in the default generation prompt. Firstly, it lists some operators selected by the heuristic search algorithm (discussed in \S~\ref{subsec:heur}). Then this example goes with the instruction ``\textit{Please generate a valid PyTorch model with selected operators}''. Next, it provides a DL model consisting of two components: model definition and input tensors (introduced in \S~\ref{subsec:bg-framework}). Finally, with such few-shot examples, \genm generates tests with the prompt template shown in Figure~\ref{fig:gen-prompt-default-2}. Placeholder \CodeIn{[LIBRARY]} in the instruction denotes the specific framework (\pt or \tf), \CodeIn{[SELECTED\_OPS]} is filled by results of heuristic search, and \CodeIn{[PLACEHOLDER...]} is the DL model to be generated by \genm.
\label{subsec:defaultgen}

\subsection{Differential Testing and Oracles}
\label{subsec:difftesting}

\parabf{Differential Testing.} Differential testing has been proven to be an effective way to detect bugs in DL frameworks~\cite{liu2023nnsmith}. The DL model generated by LLM is executed on different backends to identify bugs. In this work, we focus on the \textit{compilation} feature of DL frameworks and design a cross-check mode on two different backends as follows.

\begin{itemize}
    \item \textit{Eager backend} refers to the DL framework that directly executes DL models without \textit{compilation}. Following the development of frameworks over the past decade, eager backend has become stable and usually serves as a baseline for comparison with other backends.

    \item \textit{Compiler backend} has been the most important feature of DL frameworks since the version 2.x era, pushing the performance of DL frameworks to new heights. As an emerging technology, the compiler backend is still in rapid development, and has become an important backend to compile DL models (especially LLMs)~\cite{whitefox}. \ptinductor and \tfxla are compiler backends for \pt and \tf, respectively.
\end{itemize}

\parabf{Oracles.} We execute generated models on these two backends respectively and design two oracles to identify bugs.

\begin{itemize}
    \item \textit{Numerical inconsistency (silent incorrectness)}. The compiled model should be \textit{semantically equivalent} to the native model (as the DL compiler only optimizes the performance of DL models). Otherwise, this implies that the compiler backend incorrectly optimizes the model, resulting in \textit{silent incorrectness}. We use $| \text{input}_i - \text{other}_i | \leq atol + rtol \times | \text{other}_i |$ to compare two output tensors, where $\text{input}_i$ denotes each element in output tensor of eager backend and $\text{other}_i$ denotes each element in output tensor of compiler backend. We use the default tolerance (\ie $atol=0.001$ and $rtol=0.001$).
    
    \item \textit{Behavior inconsistency (program crashes)}. In addition to numerical inconsistency, aligning the behavior between the eager backend and the compiler backend is also very crucial. Both eager backend and compiler backend should behave the same on the same test inputs (\ie both throw errors or pass tests). Otherwise, we consider this a potential bug (\ie one throws errors, and the other one does not). 
    
\end{itemize}

 Execution and cross-checking on the two backends can be classified into three distinct states (Figure~\ref{fig:motivation1}): \textit{(i)} Cross-checking results violate the oracles, indicating a potential bug in DL frameworks, then we collect bug reports. \textit{(ii)} If a model crashes on both backends, it is an \textit{invalid} model, and we collect exception logs. \textit{(iii)} Only when execution results of both backends do not violate oracles or crash, it is a \textit{bug-free} and \textit{valid} model, and code coverage is collected.


\subsection{Prompting LLM for Analysis Summary}
\label{subsec:feedback}

\begin{figure}[htbp]
	\centering
	\includegraphics[width=0.85\columnwidth]{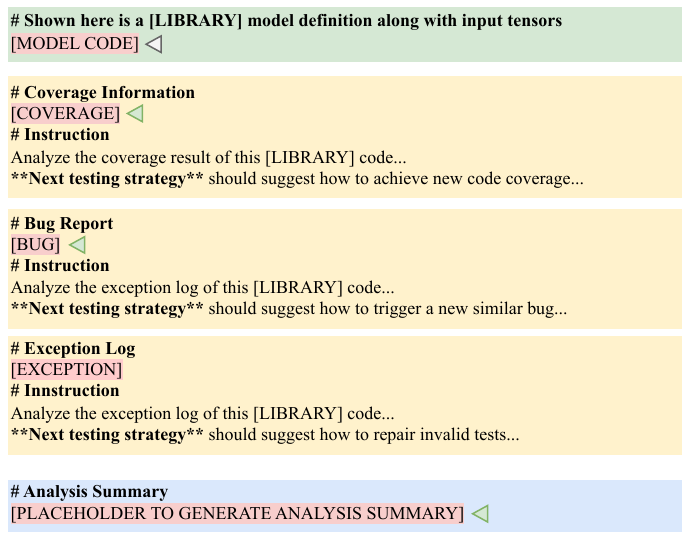}
	\caption{Analysis prompt templates of \tactic}
	\label{fig:als-prompt}
\end{figure}  

As discussed above, test execution outcomes can be classified into three distinct states, and the post-execution phase following differential testing yields semantically rich feedback information. Our core insight lies in recognizing the potential of this semantic corpus (\eg coverage data) as a valuable resource for LLMs. However, directly feeding the raw feedback into the generation LLM presents two critical challenges: \textit{(i)} \textbf{Lengthy and redundant feedback information.} The inherent redundancy of code coverage contexts (typically spanning thousands of code blocks), coupled with verbose exception logs, imposes tremendous computational and understanding burdens on \genm. \textit{(ii)} \textbf{Lack of strategic guidance}. Unprocessed feedback fails to provide \textit{explicit} and \textit{concise} instructions for subsequent iterations, leaving \genm without guidance for generation.

To address the above challenges, we introduce a \alsm to provide \textbf{clear}, \textbf{explicit}, and \textbf{concise} \textit{analysis summary} for \genm. \alsm infers \textit{explanation}, \textit{reasons}, and \textit{next testing strategy} from feedback step by step. Figure~\ref{fig:als-prompt} shows the analysis prompt template of \tactic, which begins with a contextual description: \textit{``Shown here is a \CodeIn{[LIBRARY]} model along with input tensors from the last iteration.''} The core components consist of \CodeIn{[MODEL CODE]}, which represents a DL model (introduced in \S~\ref{subsec:bg-framework}).
Next, \alsm selects different templates based on different types of feedback, and the feedback information will be subsequently filled into the corresponding placeholders (\ie \CodeIn{[COVERAGE]}, \CodeIn{[BUG]}, and \CodeIn{[EXCEPTION]}). Immediately following, the instruction makes the \alsm analyze the feedback information semantically. Finally, \alsm infers \textit{explanation}, \textit{reasons}, and \textit{next testing strategy} in \CodeIn{[PLACEHOLDER...]}. This progressive step provides explicit guidance for test generation in subsequent iterations.
More specific examples of analysis prompts will be discussed in \S~\ref{subsec:rq2}.

\subsection{Heuristic Search for Operator Selection}
\label{subsec:heur}

While LLM-based approaches can benefit from sufficient analysis summaries, they tend to use some \textit{common} and \textit{well-tested} operators (\eg Conv2d, ReLU, and Linear) for test generation~\cite{fuzzgpt}, which significantly limits model diversity (discussed in Limitation \#4). This limitation introduces the implementation of providing additional context information for \genm, a logical and natural enhancement that prioritizes less prevalent operators in test generation.

Unfortunately, operator selection is a challenging \textit{NP-hard} problem, stemming from the \textit{combinatorial explosion} inherent in DL model construction: While DL frameworks typically offer a finite operator set (numbering in the thousands), the unbounded combinations of these operators create an intractably large retrieval space. 
Notably, we identify that feedback signals can serve as indicators of intrinsic value, enabling the iterative optimization of operator selection through value-guided search.
The idea behind employing heuristic search lies in \textbf{reformulating the \revise{operator selection} task as a combinatorial optimization problem within a constrained space}: At each iteration, the algorithm selects an optimal subset of operators from the pre-collected API documents that maximizes a predefined value function. This value function, dynamically determined by accumulated feedback signals, quantifies the strategic importance of operators for subsequent iterative generation.

Inspired by the above idea, we adopt simulated annealing~\cite{van1987simulated} for operator selection in \tactic, a balance between overhead and performance. High-performance search algorithms, such as Monte Carlo Tree Search, incur prohibitive temporal overhead during incremental tree construction in iterative fuzzing cycles. Conversely, low-overhead search algorithms (\eg random sampling) fail to identify \textit{high-value} operators, compromising search performance. Simulated annealing emerges as an offline planning algorithm that efficiently approximates global optima (regulated through cooling rate $\gamma$) within constrained time budgets. For operator valuation, we design three metrics of operators: \textit{(i) Number of uses}; \textit{(ii) Exception-triggering count}; \textit{(iii) Number of new triggered line coverage}. The valuation function follows an inverse relationship with the first two factors (\ie frequency and exceptions), reflecting our emphasis on exploring \textit{underused} operators. Conversely, it maintains direct proportionality to new code coverage, prioritizing operators that can trigger new code coverage. This strategy effectively prevents the fuzzer from being trapped in coverage plateaus caused by over-testing common operators. We define the final valuation function as follows: $V_{\alpha,\beta}(x, y, z)=\frac{\alpha}{\alpha+x} + \frac{\alpha}{e^y} + \beta - e^{-\frac{z}{100}}$, where $x$, $y$, and $z$ denote for \textit{number of use}, \textit{exception-triggering count}, and \textit{number of new triggered line coverage}, respectively. The hyper-parameters $\alpha$ and $\beta$ are both initialized at $0.5$. This configuration ensures algorithmic stability before the fuzzing campaign, where $x$, $y$, and $z$ remain zero-initialized. Under this initialization setting, the value for all operators reduces to: $V_{\alpha,\beta}(0,0,0)= \alpha+\beta$. Substituting the predefined hyperparameters yields a uniform initial value $V_{0.5,0.5}(0,0,0)=1.0$, ensuring fair operator prioritization at the beginning of fuzzing.

Algorithm~\ref{algo:saos} presents the details of our \textbf{Feedback-Aware Simulated Annealing (\algo)}. The algorithm begins by verifying whether an exception was triggered in the previous iteration (Line~\ref{algo:bugflag}). If true, \algo returns the previously selected operator sequence to provide a repair opportunity for \tactic (Lines~\ref{algo:bugflag-strat}-\ref{algo:bugflag-end}). Subsequently, operator values are updated through the integration of three distinct feedback signals (Lines~\ref{algo:opvalue-start}-\ref{algo:opvalue-end}). The optimization process employs a simulated annealing framework that returns the selected operators (Lines~\ref{algo:getops-start}-\ref{algo:getops-end}), initialized with standard temperature parameters and cooling schedules (Lines~\ref{algo:init-strat}-\ref{algo:init-end}). During each annealing cycle (Line~\ref{algo:externalcycle}), $N_s$ iterations are performed to evaluate candidate operator sequences $S_{new}=[op_1,op_2,...,op_k]$ (Lines~\ref{algo:internalcycle-start}-\ref{algo:internalcycle-end}), where the composite fitness value function $F$ is calculated as the average of individual operator values ($F = \frac{1}{k} \sum_{i=1}^{k} V_{\alpha,\beta}(x_i, y_i, z_i)$). While sequences with $\Delta F \ge 0$ are typically accepted, the Metropolis criterion probabilistically permits inferior solutions (with probability $e^{\Delta F/T}$) to escape local optima (Lines~\ref{algo:value4update-strat}-\ref{algo:value4update-end}). The temperature parameter undergoes reduction according to a predefined cooling schedule until reaching termination thresholds (Lines~\ref{algo:paramupdate-strat}-\ref{algo:paramupdate-end}). Finally, \algo returns a selected operator sequence in the current fuzzing iteration (Line~\ref{algo:bestseq}).

\begin{algorithm}[t]
\caption{Feedback-Aware Simulated Annealing}
\label{algo:saos}
\DontPrintSemicolon
\SetKwProg{Fn}{Function}{:}{}

\SetKwData{deltacov}{$\Delta$Coverage}
\SetKwData{expflag}{exceptionFlag}
\SetKwData{lastops}{lastOps}
\SetKwData{op}{Op}
\SetKwData{selectedop}{selectedOps}
\SetKwData{deltaf}{$\Delta$F}
\SetKwData{opset}{opSet}
\SetKwData{isnotfirst}{isNotFirstIteration}

\SetKwFunction{opselection}{OpsSelection}
\SetKwFunction{sa}{SimulatedAnnealing}
\SetKwFunction{rn}{RandomNumber}
\SetKwFunction{rs}{RandomSelect}
\SetKwFunction{f}{F}
\SetKwFunction{rr}{RandomReal}

\SetKwInOut{Input}{Input}
\SetKwInOut{Output}{Output}

\Fn{\opselection{\deltacov, \expflag, \lastops, \opset}}{
    \Input{\deltacov, \expflag, \lastops, \opset}
    \Output{Selected operator sequence: \selectedop}
    \If{\expflag $== 1$}{   \label{algo:bugflag} 
        \selectedop $\leftarrow$ \lastops\\ \label{algo:bugflag-strat}
        \Return{\selectedop}  \label{algo:bugflag-end} 
    } 
    \If{\isnotfirst}{ \label{algo:opvalue-start}
        \For{$\op \in \lastops$} 
        {
            $\op_{used\_times} \leftarrow \op_{used\_times} + 1$\\ 
            $\op_{exp\_count} \leftarrow \op_{exp\_count} +$ \expflag\\ 
            $\op_{cov\_count} \leftarrow \op_{cov\_count} +$ \deltacov 
        }
    } \label{algo:opvalue-end}
     
    \selectedop $\leftarrow$ \sa{\opset}\\ \label{algo:getops-start}
    \Return{\selectedop} \label{algo:getops-end}
}

\Fn{\sa{\opset}}{
    $T, N_s, \gamma$  $\leftarrow$ $100, 10, 0.99$\\ \label{algo:init-strat}
    $k$  $\leftarrow$ \rn{$1,3$} \\
    $S_{cur}$ $\leftarrow$ \rs{$\opset,k$}\\ \label{algo:init-end}
    \While{T $\geq$ $T_{min}$}{ \label{algo:externalcycle}
        $n_s$ $\leftarrow$ $0$\\ \label{algo:internalcycle-start}
        \While{$n_s$ < $N_s$}{
            $S_{new}$ $\leftarrow$ \rs{$\opset,k$}\\ 
            \deltaf $\leftarrow$ \f{$S_{new}$} - \f{$S_{cur}$}\\ \label{algo:internalcycle-end}
            \If{\deltaf $\geq 0$ $|$ \rr{$0,1$} < $e^{{\deltaf/T}}$}{ \label{algo:value4update-strat}
                    $S_{cur}$ $\leftarrow$ $S_{new}$  \label{algo:value4update-end}
            }  
            $n_s \leftarrow n_s + 1$\\  \label{algo:paramupdate-strat}
        }
        $T \leftarrow T \cdot \gamma$  \label{algo:paramupdate-end}
    }
    \Return{$S_{cur}$}  \label{algo:bestseq}
}
\end{algorithm}

\subsection{Feedback-guided Prompting for Test Generation}
\label{subsec:guidance}

\begin{figure}[htbp]
	\centering
	\includegraphics[width=0.85\columnwidth]{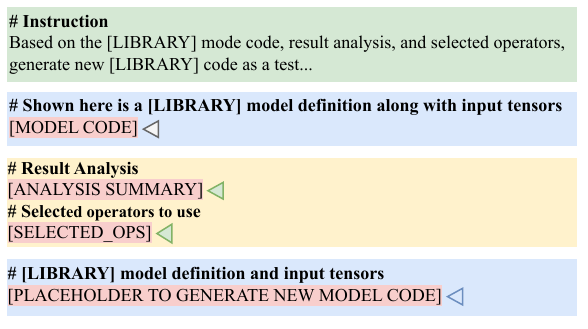}
	\caption{Feedback-guided prompt template of \tactic}
	\label{fig:gen-prompt-feedback}
\end{figure}

Feedback guidance consists of the analysis summaries generated by \alsm and selected operators searched by \algo. \tactic employs three testing strategies (\ie explore new coverage, trigger similar bugs, and repair invalid tests). As illustrated in Figure~\ref{fig:gen-prompt-feedback}, feedback-guided prompts first specify generation instructions and the DL model generated in the last iteration, followed by the interaction of \CodeIn{[ANALYSIS SUMMARY]} containing \textit{next testing strategy}. Meanwhile, \CodeIn{[SELECTED\_OPS]} proposes selected operators. The \CodeIn{[PLACEHOLDER...]} serves as the target position for the new DL model generated by \genm. 
Notably, feedback-guided prompts operate mutually exclusively with the default prompt described in \S~\ref{subsec:defaultgen}. Activation conditions of feedback-guided prompts require the simultaneous satisfaction of two criteria: \textit{(i)} the current iteration is not the first iteration. \textit{(ii)} No consecutive invalid executions have been detected in recent cycles. This conditional substitution mechanism ensures adaptive utilization of historical feedback while maintaining generation stability, with the feedback-guided prompt dynamically replacing the default template. This design enables \tactic to effectively balance exploratory test generation with feedback-driven refinement through its context-sensitive prompt switching strategy.

%% file: 4implementation.tex
\section{Implementation}
\label{sec:imple}



\parabf{Operator collection from API document.}
As discussed in \S~\ref{sec:approach}, an \textit{operator set} is needed for \algo. In this work, we collect core operators from two DL frameworks. Specifically, we implement an HTML crawler to collect operators from PyTorch with API prefixes \CodeIn{torch}, \CodeIn{torch.nn}, and \CodeIn{torch.nn.functional} as they are core operators~\cite{ptop} in PyTorch 2.x. For TensorFlow, we collect operators that XLA supports from its document~\cite{tfxlaops} as we focus on the compilation feature of the DL framework in this work. Although we currently only consider the most important subset of all operators, we can easily modify the operator set to suit specific test scenarios (\eg considering gradient-related operators to test the gradient feature of DL frameworks). 

\parabf{Generation and analysis LLMs.}
While closed-source models (\eg GPT-4o~\cite{gpt-4o}) demonstrate amazing performance, their costs are substantial through the API service. In contrast, open-source models enable local deployment but typically exhibit comparatively inferior performance. In our approach, we use DeepSeek-V3~\cite{deepseekai2024deepseekv3} as the \alsm for feedback analysis, leveraging its demonstrated superiority in most benchmarks related to text summary. Meanwhile, we select Qwen-2.5-Coder-32B~\cite{hui2024qwen2.5-coder} as \genm for test generation due to its leading performance on code generation benchmarks~\cite{evalplus}. The temperature of the Qwen model is set to one that aligns with the mechanism of fuzzing (\ie generating random and different test cases each time). This model selection strategy carefully balances the trade-off between performance and costs. Specifically: \textit{(i)} For \textit{local deployment}, we use vLLM~\cite{vllm}, an efficient LLM inference engine with optimized memory management, to host open-source local models; \textit{(ii)} For \textit{API services}, we adopt a Python package openai~\cite{openai-python} that maintains compatibility with mainstream model API services. Notably, our artifact supports flexible integration of diverse LLMs through different invocation ways (\ie local deployment and API services), allowing dynamic selection between generation and analysis models based on specific requirements and budgets. 

\revise{
\parabf{In-Context Learning Prompting.}
Prompt engineering plays a pivotal role in LLM performance~\cite{sahoo2024promptsurvey}. Since \tactic integrates two distinct LLM agents (\alsm and \genm), we design specialized prompts for each, employing lightweight few-shot In-Context Learning (ICL)~\cite{brown2020gpt3}. To balance manual effort with system extensibility, we curate a concise set of few-shot examples that comprehensively cover the spectrum of feedback categories defined in our design: \textit{(i) Coverage Information}, distinguishing whether triggering new coverage; \textit{(ii) Bug Reports}, encompassing both single-backend crashes and numerical inconsistencies; \textit{(iii) Exception Logs}, representing invalid executions where crashes occur across both backends. To construct high-quality examples, we collected representative DL programs from framework documentation and historical GitHub issues. These programs were executed to capture real-world <program, feedback> pairs. We then manually crafted the corresponding \CodeIn{[ANALYSIS SUMMARY]} to form ground-truth triplets. These triplets serve as few-shot examples to teach \alsm how to distill feedback, and subsequently guide \genm in interpreting these summaries for test generation. Although we opt for this lightweight prompting strategy over complex paradigms (e.g., Chain-of-Thought~\cite{wei2022cot}) for efficiency, it demonstrates sufficient effectiveness in our evaluation. Notably, such implementation significantly reduces the manual burden compared to traditional fuzzing techniques~\cite{liu2023nnsmith,wang2020lemon}.
}

%% file: 5evaluation.tex
\section{Evaluation}
\label{sec:eval}

\revise{
\subsection{Research Questions}
We design the following research questions (RQs) in our evaluation:
\begin{itemize}
    \item \textbf{RQ1:} Can \tactic outperform the state-of-the-art (SOTA) baselines on coverage and bug detection?
    \item \textbf{RQ2:} Can \tactic detect real-world new bugs? 
    \item \textbf{RQ3:} How does each component contribute to \tactic?
\end{itemize}
}

\subsection{Experimental Setup}

\parabf{Systems under test.}
In this work, we focus on two most popular DL frameworks (\pt~\cite{PyTorch} and \tf~\cite{Tensorflow}). The statistics of two DL frameworks are shown in Table~\ref{tab:statistic}. \revise{In the following tables, we use PT and TF to refer to \pt and \tf, respectively.}


\begin{table}[htbp]
\centering
\caption{Statistics of the SUTs}
\label{tab:statistic}
\begin{tabular}{c|ccc}\toprule
\textbf{SUT} & \textbf{Lines of Code} & \textbf{Test Language} & \textbf{Tested Version} \\\midrule
PT & 322,839  & Python  & Nightly20250214 \\
TF & 285,406  & Python  & Nightly20250228 \\
\bottomrule
\end{tabular}
\end{table}

\parabf{Fuzzing configuration and environment.}
We run \tactic on 1,000 iterations to generate tests, as a trade-off between high costs and the number of sufficient generated tests. Our experiments were conducted on Ubuntu 20.04 LTS with 4 NVIDIA V100 GPUs.

\revise{
\parabf{Baselines}. We select SOTA DL framework testing baselines from two categories (LLM-based fuzzers~\cite{titanfuzz,whitefox} and traditional fuzzers~\cite{liu2023nnsmith,liu2023neuri}). To ensure a fair comparison, we use the same oracle and differential testing (as detailed in \S~\ref{subsec:difftesting}) for baselines. We introduce the characteristics of these baselines as follows.
\begin{itemize}
    \item \titanfuzz~\cite{titanfuzz} is the first technique to directly leverage LLMs for fuzzing DL libraries, capitalizing on their implicit knowledge of language syntax and API constraints. It employs a generative LLM to produce seed programs and an infilling LLM to populate valid API parameters. Finally, it executes generated APIs and compares outputs across different hardware to detect bugs.
    \item \whitefox~\cite{whitefox} is an LLM-based white-box fuzzer that leverages compiler optimization source code to guide test generation. It uses an analysis LLM to extract optimization patterns as requirements, which are then used by a generation LLM to synthesize targeted test cases.
    \item \nnsmith~\cite{liu2023nnsmith} targets DL compilers by utilizing an SMT solver with manually defined operator specifications to generate diverse and valid computation graphs. Then it further employs a gradient-based search to find inputs that prevent floating-point exceptions, ensuring valid comparisons during differential testing.
    \item \neuri~\cite{liu2023neuri} tackles the challenge of defining complex operator constraints by employing inductive program synthesis. It infers these constraints automatically from collected valid and invalid API traces and subsequently utilizes a hybrid model generation strategy (combining symbolic and concrete execution) to synthesize diverse and valid DL models capable of supporting hundreds of operator types.
\end{itemize}
}

\parabf{\tactic variants}. We design five variants for ablation studies.

\begin{itemize}
    \item \woals discards \alsm not to generate analysis summary. \genm in \woals is prompted to generate tests based on the feedback information.
    \item \wosa generates tests without \algo, meaning that $\mathcal{M_{G}}$ generates tests only based on analysis summaries.
    \item \wocov is a variant where the coverage feedback is disabled. When the test passes, \wocov generates tests randomly with the default prompt.
    \item \wobug is a variant where the bug feedback is disabled. When violating oracles, \wobug generates tests with the default prompt template.
    \item \woexc generates random tests when an invalid test is executed. Notably, in such a situation, the program self-repair strategy is disabled.
\end{itemize}

\parabf{Metrics}. We consider two types of metrics. \textit{(i) Code coverage}: We use a Python library (namely \CodeIn{Coverage.py}~\cite{coverage-py}) to collect line code coverage. \textit{(ii) Number of detected bugs}: Notice that we also record the number of detected bugs. We report our detected bugs to the corresponding repository issues on GitHub. We record the issue as ``Confirmed'' only when the developer can reproduce it and as ``Fixed'' when a pull request (PR) linked to the issue is merged into the main branch.

\subsection{RQ1: Comparison with SOTA baselines}
\subsubsection{Default fuzzing configuration}
\begin{figure}[tbp]
    \centering
    \begin{subfigure}[b]{0.45\linewidth} %
        \includegraphics[width=\linewidth]{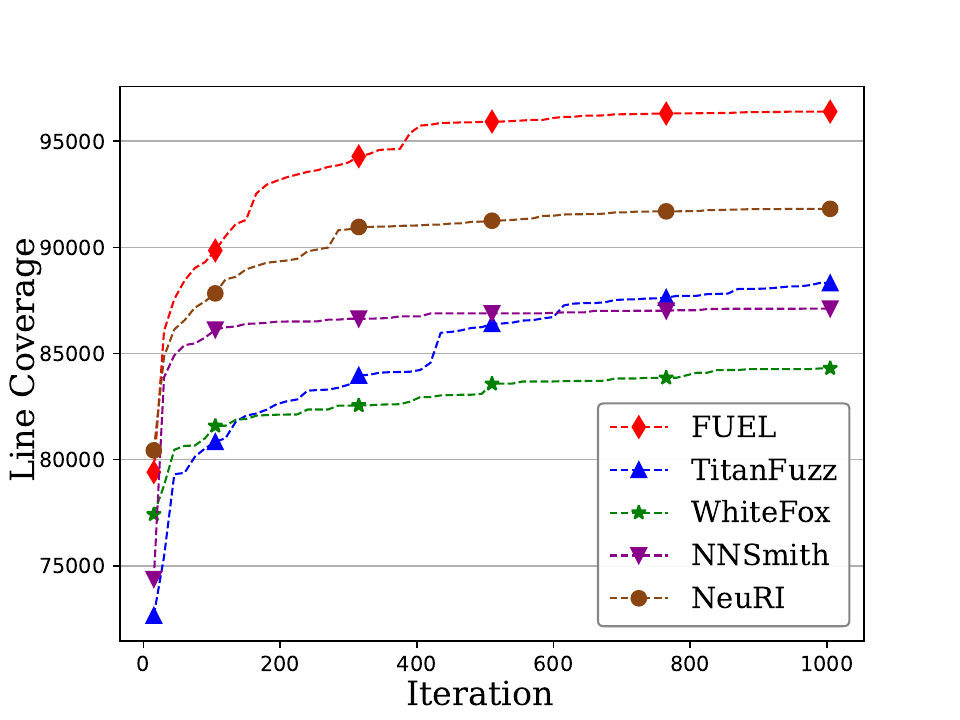}
        \caption{\pt}
        \label{pt_coverage}
    \end{subfigure}
    \hfill %
    \begin{subfigure}[b]{0.45\linewidth} %
        \includegraphics[width=\linewidth]{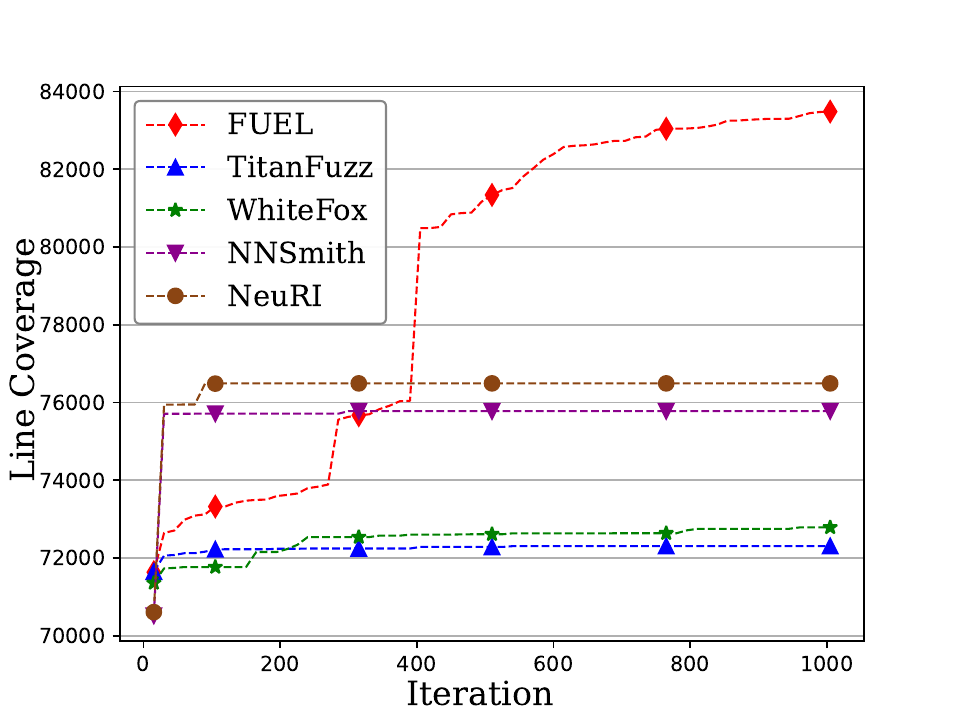}
        \caption{\tf}
        \label{tf_coverage}
    \end{subfigure}
    \caption{Code coverage comparison with SOTA baselines.}
    \label{fig:coverage}
\end{figure}

\revise{Figure~\ref{fig:coverage} demonstrates the performance of \tactic against four SOTA baselines, on both \pt and \tf. Over 1,000 iterations, \tactic achieves superior code coverage compared to all baselines, outperforming the second-best baseline \neuri by 4.48\% and 9.14\% on \pt and \tf, respectively. 
Among the baselines, \neuri demonstrates the most competitive performance. Although \neuri is also a model-level fuzzer capable of synthesizing valid models via inductive rule inference, it is fundamentally limited by its reliance on \textit{random} operator combinations. This stochastic nature prevents it from systematically exploring the rich and deep interaction space between operators. In contrast, \tactic utilizes feedback-driven guidance and heuristic search to strategically combine operators, thereby unlocking harder-to-reach code paths that random combinations miss.
Other baselines show distinct limitations. For instance, \whitefox exhibits relatively lower coverage, potentially attributed to its optimization-triggering pattern specialization. While effective for testing specific compiler optimizations, this feature inadvertently restricts model diversity through excessive template matching. These experimental results support our hypothesis that utilizing coverage information to guide test generation is more effective than LLM-based fuzzers and traditional fuzzers, achieving greater coverage across 1,000 iterations.
}
\label{subsec:rq1}

\subsubsection{24-hour fuzzing configuration}
\label{subsec:rq1-2}
\revise{
Following prior studies~\cite{xia2024fuzz4all,whitefox}, we evaluate \tactic over a 24-hour fuzzing campaign. Table~\ref{tab:rq1} reveals that \tactic generates the fewest tests (approximately 8.80\% of \titanfuzz's tests) on \pt. This phenomenon is attributed to the computationally intensive nature of the multi-agent architecture within our fuzzing loop. Despite the significant overhead, \tactic shows competitive code coverage compared to \titanfuzz and, more importantly, superior bug detection capabilities. Notably, \titanfuzz failed to detect any new bugs on the latest versions of \pt and \tf. Similarly, \neuri generates the most tests (588,610 on \pt and 649,924 on \tf) by leveraging efficient inductive rule inference and its own engineering optimization. However, its reliance on random operator combinations tends to produce numerous simple or repetitive tests. Consequently, \neuri detects fewer bugs and achieves lower code coverage than FUEL. These results indicate that in DL framework testing, simple and extensive test generation (\ie generating a single API or random operator combination) provides small rewards, especially in bug detection, emphasizing the need for strategic generation of architecturally diverse models rather than quantity-driven approaches. 
In terms of model validity and diversity, the near-perfect validity of \titanfuzz stems from its API-level fuzzing paradigm that inherently bypasses the complex constraint verification required for operator interactions. Among model-level testing approaches, \nnsmith achieves 100\% validity rate by design as it utilizes SMT solvers to construct computation graphs. However, it relies on a small set of manually defined operators that are thoroughly tested, which restricts its ability to explore more unusual operators. \tactic demonstrates significant validity improvements over \whitefox, achieving 24.82\% and 41.63\% higher validity rates in \pt and \tf, respectively. This performance gap highlights \tactic's enhanced capability in repairing invalid models through its self-repair strategy. The 24-hour experimental results position \tactic as the balanced solution that addresses the trade-off between model validity and diversity in DL framework fuzzing.
}
\begin{table}[htbp]
\centering
\caption{Comparison with baselines over 24-hour fuzzing}.\label{tab:rq1}
\resizebox{\columnwidth}{!}{%
\begin{tabular}{c|r|r|r|r|r}\toprule
\textbf{SUT} &\textbf{Technique} &\textbf{\# Bugs} &\textbf{Coverage} &\textbf{\# Valid Tests (\%)} &\textbf{\# Tests} \\\midrule

\multirow{5}{*}{PT}
& \tactic  &\textbf{8} &97,195 &2,808 (78.28\%)   &3,587 \\
& \titanfuzz &0 &\textbf{98,487} &36,734 (90.21\%)  &40,721 \\
&\whitefox &1 &88,390 &3,944 (51.55\%)  &7,651  \\
&\nnsmith &1 &87,922 & \textbf{103,248 (100.00\%)}  &103,248 \\
&\neuri &2 &92,274 & 588,610 (93.09\%)  &\textbf{632,310} \\\hline

\multirow{5}{*}{TF} 
& \tactic  &\textbf{2} &94,114 &1,930 (64.03\%)  &3,014\\
&\titanfuzz &0 &\textbf{99,778} &33,896 (93.84\%)  &36,121 \\
&\whitefox &0 &74,785 & 1,353 (22.40\%)  &6,041 \\
&\nnsmith &0 &75,777 &\textbf{47,321 (100.00\%)}  &47,321 \\
&\neuri &0 &76,493 & 649,924 (89.72\%)  &\textbf{724,392} \\
\bottomrule
\end{tabular}
}
\end{table}

\subsection{RQ2: Bug Study}
\label{subsec:rq2}

Table \ref{tab:rq2} presents the bug detection results of \tactic. We first analyze some representative detected bugs and then analyze the advantage of \tactic to trigger similar new bugs. Lastly, we explain why some detected bugs won't be fixed by DL framework developers.

\begin{table}[htbp]
\centering
\caption{Summary of our detected bugs}\label{tab:rq2}

\begin{tabularx}{\linewidth}{c|c|c|c|c}
\toprule
\textbf{SUT} & \textbf{Total} & 
\textbf{Confirmed (Fixed)} &
\textbf{Pending} & \textbf{Rejected} \\
\midrule
PT & 98 & 87 (53) & 6 & 5 \\
TF & 6 & 6 (0) & 0 & 0 \\
\hline
\textbf{Total} & 104 & 93 (53) & 6 & 5 \\
\bottomrule
\end{tabularx}
\end{table}

\subsubsection{Analysis of bug examples}
As discussed in \S~\ref{subsec:difftesting}, we use two oracles to identify bugs on two backends, including numerical inconsistency and behavior inconsistency. We introduce two representative bugs from \pt and \tf as follows.

\begin{figure}[htbp]
    \centering
    \captionsetup[subfigure]{justification=centering, aboveskip=3pt, belowskip=2pt}
    \begin{subfigure}{\linewidth}
        \centering
        \includegraphics[width=0.85\linewidth]{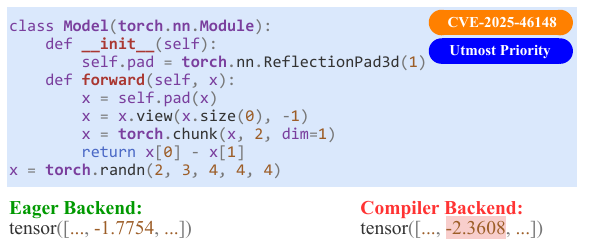}
        \caption{A \pt numerical inconsistency bug detected by \tactic}
        \label{fig:pt-example-bug}
    \end{subfigure}

    \begin{subfigure}{\linewidth}
        \centering
        \includegraphics[width=0.85\linewidth]{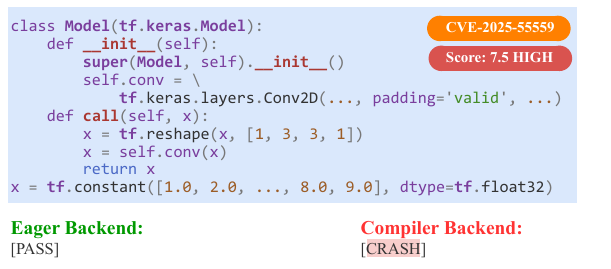}
        \caption{A \tf behavior inconsistency bug detected by \tactic}
        \label{fig:tf-example-bug}
    \end{subfigure}
    \captionsetup{aboveskip=5pt}
    \caption{Representative bugs detected by \tactic}
    \label{fig:bug-examples}
\end{figure}

Figure~\ref{fig:pt-example-bug} shows a \textbf{numerical inconsistency} bug of \pt which implements an \textbf{incorrect graph-level optimization}. 
The two tensors returned by \CodeIn{torch.chunk} are subtracted, resulting in different outputs on the two backends. The root cause of this bug is that \ptinductor attempts to fuse \CodeIn{torch.nn.ReflectionPad3d}, \CodeIn{view}, and \CodeIn{torch.chunk} in the computational graph (this optimization is also called \textit{operator fusion}~\cite{alwani2016fused}). However, in the final step of indexing (\ie \CodeIn{x[0]- x[1]}), \ptinductor incorrectly assumed that the value of the tensor must be positive (while it might be zero due to the modulo operation), causing the compiler backend to output incorrect results. This bug was labeled as \colorbox{myblue}{utmost priority} by \pt community and assigned \textbf{CVE-2025-46148}, which was fixed by a PR immediately.

Figure~\ref{fig:tf-example-bug} demonstrates a \textbf{behavior inconsistency} bug of \tf, which is a \textbf{security vulnerability} detected by \tactic. The buggy model consists of \CodeIn{tf.reshape} and \CodeIn{tf.keras.layers.Conv2D}. When setting \CodeIn{padding=\textquotesingle valid\textquotesingle} for \CodeIn{Conv2D}, the eager backend executes the model successfully, while the compiler backend throws \CodeIn{negative dimension size} error. This vulnerability can be leveraged by requiring the \CodeIn{Conv2D} parameter in \CodeIn{padding=\textquotesingle valid\textquotesingle} format, resulting in DoS attacks, which was assigned \textbf{CVE-2025-55559}. CVSS-3.x scored the severity of this vulnerability as \colorbox{myred}{"7.5 HIGH"}.

\begin{figure*}[htbp]
    \centering
    \captionsetup[subfigure]{justification=centering, aboveskip=3pt, belowskip=2pt}
    \begin{subfigure}{\linewidth}
        \centering
        \includegraphics[width=\linewidth]{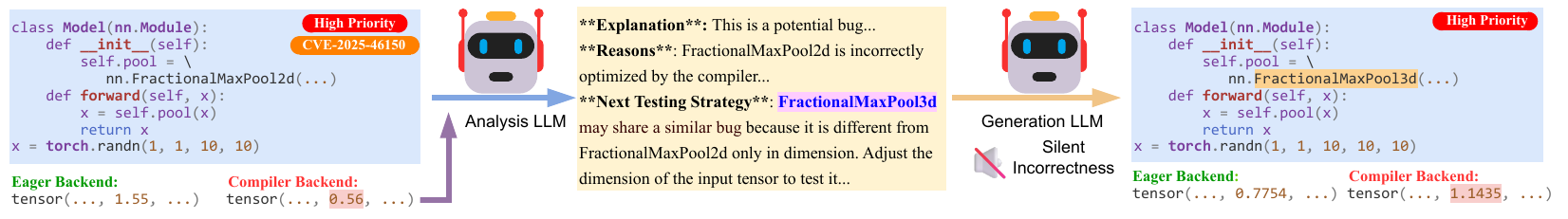}
        \caption{Trigger a new similar bug through operator overloading }
        \label{fig:similar-example-bug-1}
    \end{subfigure}

    \begin{subfigure}{\linewidth}
        \centering
        \includegraphics[width=\linewidth]{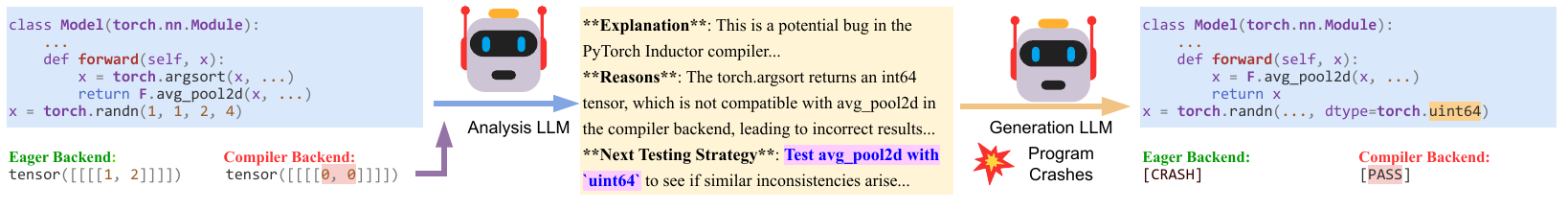}
        \caption{Trigger a new similar bug through parameter mutation}
        \label{fig:similar-example-bug-2}
    \end{subfigure}

    \begin{subfigure}{\linewidth}
        \centering
        \includegraphics[width=\linewidth]{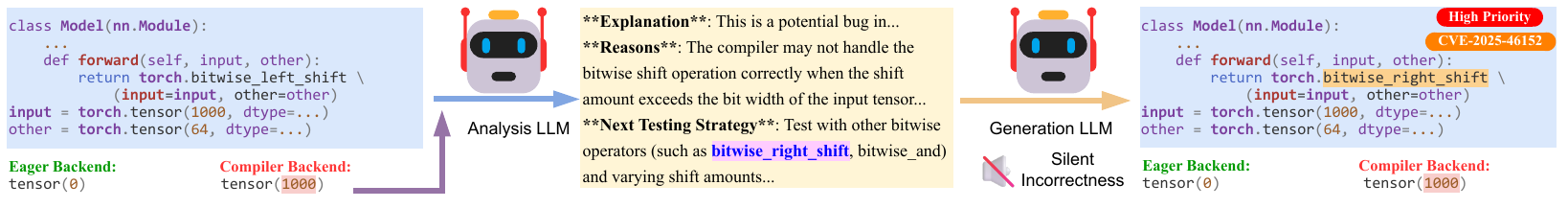}
        \caption{Trigger a new similar bug through inverse operators}
        \label{fig:similiar-example-bug-3}
    \end{subfigure}
    
    \captionsetup{aboveskip=5pt}
    \caption{The process of \tactic triggering new similar bugs with the last triggered bugs}
	\label{fig:similar-bugs}
    
\end{figure*}

\subsubsection{Analysis of detecting similar new bugs} In our approach, one of the feedback goals is to \textit{detect new bugs with the last triggered bug}. The insight behind this goal is that \textit{bug-inducing inputs are usually clustered together}~\cite{dickinson2001pursuingfailure}. \alsm can analyze current bugs and guide \genm to generate similar bug-inducing inputs in the next iteration. \ignore{Over the xxx bugs we detected, xxx bugs were found in successive iterations.} \tactic can generate similar bug-inducing inputs from three perspectives, including \textbf{operator overloading}, \textbf{parameter mutation}, and \textbf{inverse operators}.

\parabf{Operator overloading}. Figure~\ref{fig:similar-example-bug-1} illustrates the process of generating bug-inducing inputs. First, the last test triggers a bug with the \CodeIn{FractionalMaxPool2d} operator, an optimization problem with inconsistent results. Then, the analysis of the current test results is given by \CodeIn{\alsm}. It highlights that this could be a common issue with the \CodeIn{FractionalMaxPool} and suggests that test generation uses the \CodeIn{FractionalMaxPool3d} operator in the next iteration. The \CodeIn{\genm} generates the test, which is executed on different backends based on the analysis summaries. The inconsistent results after execution indicate that \CodeIn{FractionalMaxPool3d} was also incorrectly optimized by PyTorch Inductor. 
We reported these two bugs to \pt and a core developer commented on them as follows:

\begin{tcolorbox}[width=1.0\linewidth, title={}]
\textit{``Thank you for the issues. Silent incorrectness is always a high priority!''}

\makebox[\linewidth][r]{\textit{\textemdash{} A \pt core developer}}
\end{tcolorbox}

\parabf{Parameter mutation}. In addition to detecting bugs of operator overloading, \tactic can also detect bugs for different parameters (\ie data types and properties) of the same operator. Figure~\ref{fig:similar-example-bug-2} shows two bugs found by \tactic. Feedback from the last iteration is a model consisting of \CodeIn{torch.argsoft} and \CodeIn{F.avg\_pool2d}, triggering numerical inconsistency. \alsm infers that return value of \CodeIn{torch.argsoft} is \CodeIn{torch.int64} which results to all-zero output of \CodeIn{F.avg\_pool2d} on inductor. Hence, \alsm suggests using other \textit{unusual} data types (\eg \CodeIn{torch.uint64}) to test \CodeIn{F.avg\_pool2d} in the next iteration. \genm generates the test based on the suggested strategy and subsequently triggers a behavior inconsistency bug. \pt developers confirmed this bug, and we sent a PR to fix it, which was quickly approved and merged into main branch.

\parabf{Inverse operators}. Different from the above two perspectives, which trigger similar bugs on the same operator, \tactic can also target different operators. As Figure~\ref{fig:similiar-example-bug-3} shows, in the last execution, \CodeIn{bitwise\_right\_shift} outputs incorrectly on the compiler backend. \alsm infers that an inverse operator (\ie \CodeIn{bitwise\_left\_shift}) may have the same bug and suggests testing this operator in the next iteration. \genm subsequently generates the DL model consisting of \CodeIn{bitwise\_left\_shift} which triggers the silent incorrectness similar to \CodeIn{bitwise\_right\_shift}. This bug was labeled as \textbf{high priority} and assigned \textbf{CVE-2025-46152}.


\subsubsection{Rejected Bugs} Among the 104 bugs we have reported, five of them won't be fixed by developers. Developers dismiss them because the numerical inconsistency is too small. They suggested increasing the tolerance threshold of accuracy. Subsequently, we did it in our next bug reports.

\subsection{RQ3: Ablation Studies}
\label{subsec:rq3}
Similar to the observations of prior studies (\fuzzgpt~\cite{fuzzgpt}, \whitefox~\cite{whitefox}), we notice that the \pt community is more active than the \tf community as \pt developers confirm and fix issues faster. Therefore, we run our ablation experiments on \pt with the default fuzzing configuration. Table~\ref{tab:ablation} shows ablation comparison results.

\begin{table}[htbp]\centering
\caption{Comparison with ablation variants}.\label{tab:ablation}
\renewcommand{\arraystretch}{1.25} 
\begin{tabular}{c|r|r|r}\toprule
\textbf{Technique} &\textbf{\# Bugs} &\textbf{Coverage} &\textbf{\# Valid Tests (\%)} \\\midrule

\multirow{1}{*}
\tactic &\textbf{6} &\textbf{96,405} &793 (79.30\%) \\
\woals &1 &92,009 &730 (73.00\%) \\
\wosa &2 &88,201 &\textbf{849 (84.90\%)}\\
\wobug &2 &95,398 &789 (78.90\%)\\
\wocov &5 &90,540 &805 (80.50\%) \\
\woexc &4 &95,807 &801 (80.10\%) \\

\bottomrule
\end{tabular}
\end{table}

\subsubsection{Contribution of analysis LLM}

As discussed in our insight (~\S~\ref{sec:intro}), we assume that LLMs exhibit challenges in directly leveraging raw feedback information for test generation. In this ablation study, \woals directly utilizes feedback information for DL model generation through simple substitution of the original analysis summary. Compared to \woals, \tactic shows a 4.78\% improvement in code coverage. The potential reason is that \woals lacks the summary mechanism (\ie \alsm) for processing verbose coverage information, limiting its capacity to explore the code coverage space comprehensively. Notably, \woals also detects fewer bugs than \tactic, supporting our hypothesis that \textit{complicated} and \textit{redundant} bug reports fail to provide sufficiently explicit guidance for \genm to trigger more latent bugs effectively. These findings highlight the critical role of \alsm in transforming raw feedback information into concise summaries, which is beneficial to test generation.

\subsubsection{Contribution of \algo}
\tactic employs a heuristic search algorithm (namely \algo) to diversify DL models generated by \genm. Experimental result shows that \tactic outperforms \wosa both on code coverage and bug detection ability. This performance advantage stems from the limitation of LLMs tend to prioritize \textit{frequently-used} operators (\eg \CodeIn{Conv2d} and \CodeIn{ReLU}) while insufficiently exploring more \textit{rarely-used} operators (\eg \CodeIn{InstanceNorm}) that potentially improve code coverage and bug detection ability. Consequently, \wosa is free from considering too many constraints of other operators, resulting in generating more valid models. Overall, equipped with \algo, \tactic can achieve higher code coverage and detect more bugs, though this comes at the cost of a slight decrease in model validity.

\subsubsection{Contribution of different feedbacks}
\label{subsubsec: abalation:feedback}
In this work, we consider three types of feedback (\ie coverage information, bug reports, and exception logs). \wobug demonstrates a competitive code coverage rate with \tactic while showing a substantial decrease in bug detection ability (four fewer bugs) and inferior model validity. We attribute this phenomenon to the lack of bug summary generated by \alsm, resulting in \wobug reducing opportunities for triggering more bugs. Interestingly, \wocov performs an inverse pattern, revealing only marginal degradation in bug detection (merely one fewer bug than \tactic) while \tactic can achieve 6.48\% more code coverage than \wocov. Notably, the orthogonal performance patterns between \wocov and \wobug empirically confirm the complementarity of these two types of feedback (\ie coverage feedback enhances code coverage while bug feedback improves bug detection). Their integration proves essential for achieving optimal effectiveness, as evidenced by the superior performance of \tactic in both metrics when simultaneously leveraging both types of feedback.
Ablation results show that \woexc (program self-repair is disabled) is superior to \wobug but inferior to \tactic in terms of both coverage and bug detection. During the whole fuzzing campaign, \tactic generates 228 invalid models, successfully repairing \textbf{121 (53.07\%)} instances. We manually checked these successful repairs, finding that 32 instances trigger previously uncovered code coverage and two instances expose new bugs. In contrast, \woexc misses these bugs and code coverage, which show the superiority of our designed program self-repair strategy.

%% file: 6discussion.tex
\section{Discussion}
\label{sec:disc}

\revise{
\subsection{Sensitivity Analysis of LLM Backends}
\label{subsec:llm-sensitivity}
To evaluate the sensitivity of \tactic on different LLM backends, we experiment with three combinations of \alsm and \genm. In addition to default implementation (\S~\ref{sec:imple}), we consider two additional LLM pairs. \textit{(i)} Qwen-2.5-Coder-32B~\cite{hui2024qwen2.5-coder} paired with Qwen3-30B~\cite{qwen3}: These two open-source LLMs were selected due to their compatibility with our local environment, and this setting allows us to examine how much performance degradation FUEL would experience in a zero-cost budget scenario. \textit{(ii)} Claude-Sonnet-4~\cite{claude-4} paired with GPT-5~\cite{gpt-5}. These LLMs were chosen based on their SOTA performance on coding benchmark~\cite{swebench}, and this setting allows us to investigate the performance gains when inspiring the maximum potential of FUEL.

Table~\ref{tab:LLM analysis} shows the comparison results across different LLM backends on the same configuration as our ablation study. Qwen-2.5-Coder-23B and Qwen3-30B combination discovered only two bugs, substantially fewer than the six found by our default configuration, and the coverage decreased by 7.43\%. Claude Sonnet 4 and GPT 5 discovered eight bugs and improved coverage by 4.66\%. However, the cost for 1,000 iterations reached \$9.63, which is significantly higher than the \$1.68 (approximately 5.7$\times$) of our default configuration. In contrast, our original implementation is a more \textbf{cost-effective} choice under constrained budgets.

\begin{table}[htbp]
\centering
\caption{Performance of \tactic across different LLM backends}\label{tab:LLM analysis}

\renewcommand{\arraystretch}{1.25}

\resizebox{\columnwidth}{!}{%
\begin{tabular}{c|c|r|r|r}
\toprule
\textbf{\genm} & \textbf{\alsm} & \textbf{\# Bugs} & \textbf{Coverage} & \textbf{\$Costs} \\
\midrule

Qwen-2.5-Coder-32B & DeepSeek-V3 & 6 & 96,405 & \$1.68 \\
Qwen-2.5-Coder-32B & Qwen3-30B   & 2 & 89,244 & \$0.00 \\
Claude-Sonnet-4    & GPT-5       & \textbf{8} & \textbf{100,894} & \textbf{\$9.63} \\
\bottomrule
\end{tabular}%
}
\end{table}

The above results indicate that \tactic is moderately sensitive to the choice of LLM backend. Lower-tier open-source LLMs tend to generate less diverse and valid tests and therefore detect fewer bugs. High-end frontier LLMs can better leverage feedback and thus improve the quality of both analysis and generation, although the improvements come at a substantially higher monetary cost. Such moderate sensitivity suggests that \tactic can maintain competitive performance even when deployed with mid-tier LLMs, while still retaining the ability to benefit from stronger LLMs when available. Notably, as future LLMs offer stronger reasoning, higher code generation, and more reliable feedback summarization ability, \tactic can directly inherit these advancements without requiring re-implementation. As a result, \tactic's performance can be further enhanced with the progress of LLMs.


\subsection{Limitations}
\tactic incurs inherent computational costs due to the use of an LLM-based multi-agent architecture within the fuzzing loop, which results in notable challenges in both time and monetary overhead. \textit{(i) Time cost} arises from the limited throughput of test generation in \tactic. As detailed in \S~\ref{subsec:rq1-2}, \alsm must process a large amount of feedback in each iteration, which restricts the number of generated tests. Our current implementations demonstrate an average latency of \textbf{24.09} seconds per iteration. Fortunately, this limitation is expected to be alleviated by the rapid evolution of LLM inference optimization techniques~\cite{vllm}. \textit{(ii) monetary cost} is introduced by the token-based billing model of commercial LLM API services, especially because feedback analysis consumes a substantial number of tokens. To reduce this expense, our implementation relies on cost-effective API providers such as the DeepSeek API~\cite{deepseek-api}. The complete experimental evaluation conducted using the DeepSeek API incurred a total cost of approximately \textbf{\$40.89}, corresponding to roughly \textbf{\$0.39 per detected bug} in our experiments.
}

%% file: 7related.tex
\section{Related Work}
\label{sec:related}

\subsection{LLM-Based Fuzzing}
\titanfuzz~\cite{titanfuzz} is one of the pioneering works in LLM-based fuzzing, which leverages two LLMs to generate and infill DL library APIs, respectively. Subsequently, \fuzzall~\cite {xia2024fuzz4all} believes LLMs have been trained on the multilingual corpus to serve as a universal fuzzer for the different SUTs with different programming languages. \sedar~\cite{fu2024sedar} leverages LLMs to generate high-quality initial seeds by adapting test cases from other Database Management Systems (DBMSs), improving code coverage in target DBMS fuzzing. LLMIF~\cite{wang2024llmif} enhances IoT protocol fuzzing by using LLMs to extract protocol information and reason about device responses from protocol specifications. ChatAFL~\cite{meng2024large} utilizes LLMs to transform human-readable protocol specifications into machine-readable formats, enabling the fuzzer to generate and mutate protocol messages systematically. PromptFuzz~\cite{promptfuzz} employs coverage-guided prompt fuzzing techniques to generate fuzz drivers iteratively. More recently, KernelGPT~\cite{kernelgpt} automates the generation of Syzkaller specifications using LLMs, leveraging their extensive pre-training on kernel code and documentation to create valid syscall sequences. Unlike previous LLM-based fuzzers, \tactic is the \textit{first} fuzzing technique to leverage an analysis LLM to analyze various semantic information in the feedback and achieve promising performance.

\subsection{DL Framework Testing}
In recent years, DL framework testing has become an active topic, and a large number of fuzzers have emerged, which can be classified into traditional fuzzers~\cite{pham2019cradle,wei2022free,xie2022docter,deng2022fuzzing,yang2023fuzzing,wang2020lemon,guo2020audee,gu2022muffin,liu2023nnsmith,liu2023neuri,mu2024devmut} and LLM-based fuzzers~\cite{titanfuzz,fuzzgpt,whitefox}. 

One of the pioneering works is CRADLE~\cite{pham2019cradle}, which detects inconsistencies by running pre-built DL models across multiple low-level backends of Keras~\cite{Keras2020}. To enhance the diversity of models, LEMON~\cite{wang2020lemon} and AUDEE~\cite{guo2020audee} build upon CRADLE by applying pre-defined mutation rules to seed models and inputs, generating a broader range of test cases. \nnsmith~\cite{liu2023nnsmith} models each DL operator with input constraints and shape inference, using SMT solving to generate valid models.  Subsequently, \neuri~\cite{liu2023neuri} is proposed to synthesize DL models via rule inference automatically. Unfortunately, the above traditional fuzzers need \textit{manual} design for mutation rules, operator specifications, or synthesis rules, which need domain-specific expert knowledge. With the success of LLMs in software testing~\cite{lemieux2023codamosa,feng2024prompting}, researchers have attempted to apply LLMs to fuzzing DL frameworks. \titanfuzz~\cite{titanfuzz} is the first work to leverage LLMs to generate API call sequences directly. Then \fuzzgpt~\cite{fuzzgpt} is proposed to generate edge-case code via in-context learning and fine-tuning. 
\whitefox~\cite{whitefox} proposes a white-box compiler fuzzing technique aided by LLMs, targeting the generation of specific inputs to trigger optimization features in compilers. However, all of them struggle to generate diverse and valid DL models, and \tactic is proposed to fill this gap by adopting a heuristic search for operator selection and a program self-repair strategy by mining the potential of feedback.

%% file: 8conclusion.tex
\section{Conclusion}
\label{sec:conc}
In this work, we propose \tactic, the first feedback-driven fuzzer for DL frameworks, which leverages LLMs to process feedback information in the fuzzing loop. \tactic employs an LLM-based multi-agent architecture comprising an analysis LLM and a generation LLM: the former infers concise analysis summaries from feedback information to guide the latter in generating tests that help to improve code coverage and bug detection. Moreover, based on multiple feedback guidance, \tactic adopts a heuristic search algorithm and program self-repair strategy to enhance model diversity and validity, respectively. Our extensive experiments validate the effectiveness of each component and demonstrate \tactic's strong capability in detecting bugs and improving code coverage compared to SOTA baselines.
